\documentclass[10pt, twocolumn, letterpaper, conference]{IEEEtran}
\usepackage{amsmath,graphicx}
\usepackage{longtable}
\usepackage{booktabs}
\usepackage{multirow}
\usepackage{amssymb}
\usepackage{amsfonts}
\usepackage{algorithm,algorithmicx,algpseudocode}
\usepackage{eso-pic}

\newcommand{\argmin}{\operatornamewithlimits{arg\, min}}

\makeatletter
\def\ps@IEEEtitlepagestyle{%
	\def\@oddfoot{\mycopyrightnotice}%
	\def\@evenfoot{}%
}
\def\mycopyrightnotice{%
	{\footnotesize  978-1-6654-8087-1/22/ \$31.00 \textcopyright 2022 IEEE\hfill}
	{}
	\gdef\mycopyrightnotice{}
}
\usepackage{eso-pic}
\newcommand\AtPageUpperMyright[2]{\AtPageUpperLeft{%
		\put(\LenToUnit{0.08\paperwidth},\LenToUnit{-1.1cm}){%
			\parbox{0.78\textwidth}{\raggedleft\fontsize{10}{11}\selectfont #1}}%
	}}%
	\newcommand{\conf}[1]{%
		\AddToShipoutPictureBG*{%
			\AtPageUpperMyright{#1}
		}
	}

\begin{document}
%\title{OFDM Sparse Channel Estimation using Sparsity Domain Smoothing Based Thresholding Recovery Method}
\title{Sparsity Domain Smoothing Based Thresholding Recovery Method for OFDM Sparse Channel Estimation}
\conf{2022 3$\rm{0^{th}}$ International Conference on Electrical Engineering (ICEE)}
%\author{Mohammad Hossein Bahonar$^\dagger$, Reza Ghaderi Zefreh$^\dagger$, 
%and Rouhollah Amiri $^\ddagger$,\\
%$^\dagger$ Department of Electrical and Computer Engineering, Isfahan University of Technology, Isfahan, Iran \\
%$^\ddagger$ Department of Electrical and Computer Engineering, Sharif University of Technology, Tehran, Iran\\
%Emails: \{ mh.bahonar@ec.iut.ac.ir, r.ghaderi@ec.iut.ac.ir, and amiri\_rouhollah@ee.sharif.edu \}
%}
\author{\IEEEauthorblockN{Mohammad Hossein Bahonar}
\IEEEauthorblockA{\textit{\footnotesize Dept. of Electrical and Computer Engineering} \\
\textit{Isfahan University of Technology}\\
Isfahan, Iran \\
Email: mh.bahonar@ec.iut.ac.ir}
\and
\IEEEauthorblockN{Reza Ghaderi Zefreh}
\IEEEauthorblockA{\textit{\footnotesize  Dept. of Electrical and Computer Engineering} \\
\textit{Isfahan University of Technology}\\
Isfahan, Iran \\
Email: r.ghaderi@ec.iut.ac.ir}
\and
\IEEEauthorblockN{Rouhollah Amiri}
\IEEEauthorblockA{\textit{\footnotesize  Dept. of Electrical and Computer Engineering} \\
\textit{Sharif University of Technology}\\
Tehran, Iran \\
Email: amiri\_rouhollah@ee.sharif.edu}}

%\author{Mohammad Hossein Bahonar$^\dagger$, ,\\
%$^\dagger$ Department of Electrical and Computer Engineering , Isfahan University of Technology, Isfahan, Iran \\
%Emails: mh.bahonar@ec.iut.ac.ir, 
%}

\maketitle

\begin{abstract}
Due to the ever increasing data rate demand of beyond 5G networks and considering the wide range of Orthogonal Frequency Division Multipllexing (OFDM) technique in cellular systems, it is critical to reduce pilot overhead of OFDM systems in order to increase data rate of such systems.
Due to sparsity of multipath channels, sparse recovery methods can be exploited to reduce pilot overhead.
OFDM pilots are utilized as random samples for channel impulse response estimation. 
We propose a three-step sparsity recovery algorithm which is based on sparsity domain smoothing.
Time domain residue computation, sparsity domain smoothing, and adaptive thresholding sparsifying are the three-steps of the proposed scheme.
To the best of our knowledge, the  proposed sparsity domain smoothing based thresholding recovery method known as SDS-IMAT has not been used for OFDM sparse channel estimation in the literature.
Pilot locations are also derived based on the minimization of the measurement matrix coherence.
Numerical results verify that the performance of the proposed scheme outperforms other existing thresholding and greedy recovery methods and has a near-optimal performance.
The effectiveness of the proposed scheme is shown in terms of mean square error and bit error rate.
\end{abstract}


\section{Introduction}
Considering the increasing number of users in cellular networks and their increasing data rate demand, it is critical to develop new technologies and network architectures in order to support the ever increasing data rate demand of beyond 5G networks.
Many concepts and technologies such as a multiple-input-multiple-output (MIMO) system \cite{RMHB_MIMO}, device-to-device (D2D) communications \cite{RMHB_D2D}, dense cellular networks \cite{RMHB_Dense}, cellular vehicle-to-everything \cite{RMHB_CV2X} communications, and intelligent reflecting surfaces \cite{RIRS} have been proposed in the literature to increase capacity of cellular networks.
On the other hand, data overhead reduction can also be considered as a tool to provide coverage to a larger number of users.
Due to the capability of an orthogonal frequency division multiplexing (OFDM) system to combat multipath fading channels by providing flat fading channels over all subcarriers
\cite{Survey_Ref25}, this technique has been widely used in communication systems. 
Hence, data overhead reduction of OFDM-based communication systems can be an important issue.

Some of the subcarriers of an OFDM-based communication system are allocated to pilots which are data overheads of such systems.
The number of pilots can be reduced under certain conditions such as existence of sparse wireless channels.
A wireless channel is usually a multipath channel and can be modeled as a sparse channel having a small number of significant paths
\cite{R003}.
The equivalent discrete-time channel will also have a sparse impulse response in time domain.
In order to estimate the sparse channels of an OFDM-based communications system, pilot-assisted sparse channel estimation techniques can be used
\cite{R004,R005}.
The proposed schemes of some researches such as \cite{Tran_4,Tran_5} are based on Compressed Sensing (CS) which states that a sparse signal can be successfully reconstructed from very few samples \cite{CS_Donoho}.
These methods estimate the sparse channel impulse response by taking into account the inherent sparsity and result in a better estimation performance in terms of mean square error (MSE) or bit error rate (BER) than conventional channel estimation methods such as Least Square (LS) and Minimum MSE (MMSE).

The performance of sparse channel estimation in an OFDM-based communication system depends on the reconstruction method and the pilot placement algorithm
\cite{Tran_4,Tran_5,Taubock_Leakage_Sparsity_Enhancing}.
In \cite{Let_Chen_Pilot_CEO}, by minimizing MMSE of the channel estimation using Cross-Entropy Optimization (CEO), a pilot placement algorithm for OFDM is proposed.
By minimizing the coherence, a random search method \cite{Tran_12} and a deterministic allocation method \cite{Let_Chenhao_Deterministic_Pilot,Pakrooh_OFDM} are proposed for pilot placement in OFDM systems.
In \cite{Pakrooh_OFDM}, it is shown that the non-uniform patterns based on Cyclic Difference Set (CDS) are optimal.
As a solution to the pilot placement problem in MIMO-OFDM systems, in 
\cite{Tran_Xueyun_MIMO}, a multiple random search and a genetic algorithm based method are suggested. Also, a random-generation-based method is proposed in \cite{Tran_4}.
Equispaced pilots are investigated in \cite{R006}. The authors consider the orthogonal matching pursuit (OMP) algorithm which is a greedy sparse reconstruction algorithm.
Thresholding-based sparse recovery methods such as Iterative Method with Adaptive Thresholding (IMAT) \cite{R001} has also been proposed in the literature as a sparse reconstruction tool.

In this paper, we investigate sparse channel estimation of OFDM-based communication systems.
Due to the superiority of thresholding-based sparse recovery methods compared to greedy approaches, we propose our three-step thresholding-based sparse recovery algorithm.
Time domain residue computation, sparsity domain smoothing, and adaptive thresholding sparsifying are the three steps of the proposed scheme, respectively.
We also investigate pilot location design based on measurement matrix coherence minimization.
We employ windowing approach as the smoothing operation in the second step of the proposed scheme.
The smoothing operation should be implemented in the sparsity domain and thus frequency domain representation of the OFDM system is also elaborated in the system model.
To the best of our knowledge, the thresholding recovery methods based of sparsity domain smoothing has not been previously employed for sparse channel estimation of OFDM systems in the literature.
Simulation results indicate that the proposed scheme outperforms other existing thresholding and greedy sparse reconstruction methods in terms of MSE and BER.
The proposed scheme is also compared with an oracle estimator and the near-optimal performance of the proposed scheme is demonstrated.

The rest of the paper is organized as follows. Section \ref{SysModel} describes the channel impulse response measurement of the OFDM-based communication system. 
Section \ref{PrpMethod} describes the proposed recovery method that utilizes sparsity domain smoothing. The pilot location design is investigated in Section \ref{PrpMethod}.
The simulation results are reported in Section \ref{SimRes}. 
Section \ref{Conc} concludes the paper. 
 
%BS CUE CL
\section{System Model}
\label{SysModel}
Consider the downlink spectrum of a cellular system where a single antenna BS transmits data to multiple CUEs.
There exist $N$ CUEs in the cell where $i$-th CUE is denoted as $c_i$.
Similar to currently utilized cellular networks, it is assumed that orthogonal CLs are used by CUEs.
Hence, interference management issue among CUEs is solved.
The OFDM technique is also used for data transmission among the BS and each CUE due to the it capability to combat multipath fading.
The channel between the BS and and $c_i$ which is denoted as $h_i$ has frequency-selective fading and its coherence time is much larger than the OFDM symbol duration.
The continuous time domain impulse response of $h_i$ can be formulated as
\begin{equation}
\label{ChImpTime}
h_{i}(t) = \sum_{k=0}^{K-1} \bar{\alpha}_{i,k} \delta (t-t_{i,k}),
\end{equation}
where $\bar{\alpha}_{i,k} \in \mathbb{C}$ is the channel coefficient of $k$-th path of the CL between the BS and $c_i$.
The time delay of the path is also denoted as $t_{i,k}$.
The equation \eqref{ChImpTime} can also be represented in a discrete form as
\begin{equation}
\label{ChImpDisc}
h_{i}[n] = \sum_{l=0}^{L-1} \alpha_{i,l} \delta[n-l],
\end{equation}
where $\alpha_{i,l} \in \mathbb{C}$ denotes the complex channel gain coefficient of the $l$-tap delay path of the CL between the BS and $c_i$.
The equation \eqref{ChImpDisc} is derived by sampling the continuous channel impulse response.
This representation that is a finite impulse response filter of length $L$ can be used in discrete and sparse signal processing applications.
It can be concluded that the channel is sparse which means the channel vector
${\bf h}_{i} = [h_{i}[0],h_{i}[1],...,h_{i}[L-1] ]^T \in \mathbb{C}^{N \times 1}$
has only $K$ nonzero elements where $K \ll L$.

It is assumed that OFDM system has a total number of $N$ subcarriers where $N_p$ of them are used for pilot symbols.
The received data at $k$-th subcarrier and $n$-th OFDM frame of $c_i$ $(1 \leq k \leq N)$ can be formulated as
\begin{equation}
r_{i}(n,k) = X_{i}(n,k) H_{i}(n,k)+V_{i}(n,k),
\label{Eq1}
\end{equation}
where $X_{i}(n,k)$, $H_{i}(n,k)$, and $V_{i}(n,k)$ are the $k$-th elements of 
${\bf \tilde{X}_{i}}(n) \in \mathbb{C}^{N\times 1}$, 
${\bf H_{i}}(n) \in \mathbb{C}^{N\times 1}$, and
${\bf V_{i}}(n) \in \mathbb{C}^{N\times 1}$ 
which are the frequency domain representations of $n$-th transmitted OFDM symbol, the multipath channel at the time of the OFDM symbol, and additive white Gaussian noise (AWGN) at the time of the OFDM symbol, respectively.
The frequency domain representations of the transmitted OFDM symbol and the multipath channels can be expressed as
\begin{align}
{{\bf \tilde{X}}_{i}}(n) &= { {\bf F}} {{\bf x}_{i}}(n),
\\
{{\bf H}_{i}}(n) &= { {\bf F}} {{\bf \bar{h}}_{i}}(n),
\end{align}
where ${{\bf x}_{i}}(n) \in \mathbb{C}^{N \times 1}$ is the $n$-th OFDM symbol that is transmitted from BS to $c_i$ and 
${\bf \bar{h}_{i}}(n) \in \mathbb{C}^{N \times 1}$ is the zero-padded discrete impulse response of the channel between the BS and $c_i$ as expressed in \eqref{ChImpDisc}.
The standard $N \times N$ DFT matrix is also denoted as ${\bf F} \in \mathbb{C}^{N \times N}$.

The transmitted data of each CUE and pilot symbols form a $N \times 1$ vector which is modulated according to the OFDM technique using Inverse Fast Fourier Transform (IFFT).
Each OFDM symbol is transmitted on $N$ subcarriers.
The pilot symbols are values of the subcarriers of the OFDM symbol which are placed at specific locations called pilot locations.
The pilot symbols of the CL between the BS and $c_i$ are $x_{i}[n],~ n \in \Lambda_{i}$.
The set of pilot locations of the CL between the BS and $c_i$ is denoted as $\Lambda_{i}$ and can be expressed as
\begin{equation}
\label{EqPilot}
\Lambda_{i}=\{ \lambda_{i,1}, \lambda_{i,2}, ... ,\lambda_{i,N_p} \},
\end{equation}
where $\lambda_{i,k}$ is the location of the $k$-th pilot of the CL between the BS and $c_i$.
Equation \eqref{EqPilot} states that the cardinality of pilot locations set for all CLs is equal to $N_p$.

Let ${\bf r}_{i}(n)  \in \mathbb{C}^{N \times 1}$ denote the vector of $n$-th OFDM symbol received samples at $c_i$.
As a result the equation \eqref{Eq1} can be expressed in a matrix form as
\begin{equation}
\label{EqRXHW}
{\bf R}_{i}(n) ={\bf X}_{i}(n) {\bf W} {\bf h}_{i}(n) + {\bf V}_{i}(n)
\end{equation}
where
${\bf X}_{i}(n) = {\rm diag}\{ \tilde{X}_{i}(n,1), \tilde{X}_{i}(n,2), ..., \tilde{X}_{i}(n,N) \} \in 
\mathbb{C}^{N \times N}$
is the diagonal matrix of the subcarriers of $n$-th OFDM symbol of $c_i$.
The $K$-sparse impulse response of the multipath channel during the transmission of $n$-th OFDM symbol is denoted as ${\bf h}_{i}(n) \in \mathbb{C}^L$.
is the $K$-sparse impulse response of the channel, and
The partial FFT matrix which includes the first $L$ columns of a standard $N \times N$ FFT matrix is denoted as ${\bf W} \in \mathbb{C}^{N \times L}$.

We aim to investigate the performance of our proposed pilot allocation method using a sparsity domain interpolation based recovery method.
As a result, the equation \eqref{EqRXHW} can be just considered at pilot positions instead of all subcarriers.
In addition to that, we solve the pilot allocation problem for all OFDM symbols.
Hence, without loss of generality, the equation \eqref{EqRXHW} can be expressed as
\begin{equation}
\label{r_XWh_MIMO_p}
{\bf R}_{i,{\Lambda_{i}}} =
{\bf X}_{i,{\Lambda_{i}}} {\bf W}_{\Lambda_{i}} {\bf h}_{i}
 + {\bf V}_{i,{\Lambda_{i}}},
\end{equation}
where
${\bf R}_{i,{\Lambda_{i}}} \in \mathbb{C}^{N_p \times 1}$  
is the vector of received pilot subcarriers at $c_i$,
${\bf W}_{\Lambda_{i}} \in \mathbb{C}^{N_p \times L}$ is the DFT submatrix with $N_p$ rows associated with the pilot subcarriers, and
${\bf V}_{i,{\Lambda_{i}}}$ is the AWGN of the associated with the pilot subcarriers in frequency domain.
Assuming equal value pilot symbols, the received data corresponding to pilot symbols denoted as 
${\bf \tilde{H}}_{i,\Lambda_{i}}$ can be expressed as
\begin{equation}
\label{EqCS}
{\bf \tilde{H}}_{i,\Lambda_{i}} = {\bf W}_{\Lambda_{i}} {\bf h}_{i}
 + {\bf V}_{i,{\Lambda_{i}}},
\end{equation}
where
${\bf H}_{i,{\Lambda_{i}}}$,
${\bf W}_{\Lambda_{i}}$, and
${\bf h}_{i}$
are the observation vector, measurement matrix, and K-sparse input vector, respectively according to the CS framework.

\section{Proposed Scheme}
\label{PrpMethod}
In this section, our sparse channel estimation approach is proposed.
In the first subsection, the proposed sparse reconstruction algorithm is discussed and the usage of sparsity domain smoothing and interpolation is investigated.
In order to evaluate the performance of the proposed scheme, OFDM pilot allocation should be investigated.
Hence, the pilot allocation problem is discussed in the second subsection by minimizing coherence of the measurement matrix.

\subsection{Sparse Reconstruction Using Sparsity Domain Interpolation}
Many sparse recovery algorithms has been used to estimate sparse multipath channels.
It has been shown that thresholding-based algorithms have better performance compared to greedy methods.
Hence, we select IMAT as our initial sparse recovery algorithm and improve its performance by adding a sparsity domain interpolation operation to it.

The IMAT algorithm is a sparsity-based random sampling recovery method.
Our main goal is to estimate sparse impulse response of channel between the BS and $c_i$ (${\bf h}_{i}$) from noisy random samples of CFR (${\tilde{\bf H}}_{i,{\Lambda_{i}}}$). 
A successive reconstruction method is utilized in the IMAT algorithm in order to perform sparse signal recovery where an adaptive threshold is used to sparsify the signal at each iteration \cite{R001}.
The inverse or pseudo-inverse of the measurement matrix is needed in the IMAT algorithm.
Therefore, the pseudo-inverse of the measurement matrix can be derived using Moore-Penrose pseudo-inverse \cite{MP_Book} as 
\begin{align}
{\bf W}_{\Lambda_{i}}^{+} &= 
{\bf W}_{\Lambda_{i}}^{H} 
{( {\bf W}_{\Lambda_{i}} {\bf W}_{\Lambda_{i}}^{H} )}^{-1} 
\nonumber
\\ 
&= 
{\bf W}_{\Lambda_{i}}^{H}  (\frac{1}{N} {\bf I}_{{N_p}\times{N_p}})
= \frac{1}{N} {\bf W}_{\Lambda_{i}}^{H},
\end{align}
where the pseudo-inverse of the measurement matrix is denoted as
${\bf W}_{\Lambda_{i}}^{+} $.

Our proposed recovery scheme consists of three steps.
To the best of our knowledge the following approach which consists of a smoothing step has not been proposed in the literature for OFDM sparse channel estimation.
The first step is to find the time domain residue using the measurement matrix pseudo-inverse as
\begin{align}
\tilde {\tilde{{\bf h}}}_{i,k}
&= \lambda {\bf W}_{\Lambda_{i}}^{+}
( {\tilde{\bf H}}_{i,{\Lambda_{i}}} - {\bf W}_{\Lambda_{i}} 
{\bf \tilde{h}}_{i,k-1} )
\nonumber
\\
&= \frac{\lambda}{N} {\bf W}_{\Lambda_{i}}^{H}
( {\tilde{\bf H}}_{i,{\Lambda_{i}}} - {\bf W}_{\Lambda_{i}} {\bf \tilde{h}}_{i,k-1} ),
\label{IMATI_Eq1}
\end{align}
where $\tilde {\tilde{{\bf h}}}_{i,k} \in \mathbb{C}^{L}$ is the time domain residue of the estimated impulse response at $(k-1)$-th iteration
and ${\bf \tilde{h}}_{i,k-1} \in \mathbb{C}^{L}$ is the sparse estimated channel impulse response at $(k-1)$-th iteration.
The second step is to use a smoothing function which is formulated as
\begin{equation}
\hat {{{{\bf h}}}}_{i,k} = 
f (\tilde {\tilde{{\bf h}}}_{i,k}) + {\bf \tilde{h}}_{i,k-1} ,
\label{IMATI_Eq2}
\end{equation}
where $\hat {{{{\bf h}}}}_{i,k}$ is the non-sparse estimated channel impulse response at $k$-th iteration and the smoothing function is denoted as $f(.): \mathbb{C}^{L} \rightarrow \mathbb{C}^{L}$.
Windowing can be used as a potential smoothing operation.
The third step is to sparsify the non-sparse $\hat {{{{\bf h}}}}_{i,k}$ using an adaptive threshold described as
\begin{equation}
\tilde{h}_{i,k}(i) = 
\left\{
\begin{matrix}
\hat {{{h}}}_{i,k}(i) & 
\vert \hat {{{h}}}_{i_k}(i) \vert > \beta {\rm e}^{-\alpha k}
\\ 
0 & {\rm otherwise}
\end{matrix}
\right.
\label{IMATI_Eq3}
\end{equation}
where $\alpha$ and $\beta$ are adaptive thresholding parameters.
The recovered signal converges to the original sparse signal when the number of iterations is sufficient and the parameters $\lambda$, $\alpha$, and $\beta$ are chosen appropriately.

\subsection {Pilot Allocation Using Measurement Matrix Coherence Minimization}
In this section, we illustrate the proposed pilot allocation method.
The performance of the proposed scheme depends on the recovery method as well as the pilot locations.
Therefore, minimization of the coherence of the measurement matrix is utilized as a metric to find the pilot locations.

The measurement matrix preserves the information from the K-sparse input vector in the observed vector. 
In order to determine a proper measurement matrix, its coherence \cite{CS_Intro} denoted by $\mu$ can be computed.
The K-sparse channel impulse response is guaranteed to be perfectly recovered when
$\mu_{{\bf W}_{\Lambda_{i}}} < \frac{1}{2K}$
\cite{Pakrooh_7}.
The pilot positions can be optimized using the coherence of each CUE which is defined as 
\begin{equation}
\label{Eq11}
\mu_{{\bf W}_{\Lambda_{i}}} = 
\max_{\underset{i \neq j}{ 1 \leq i,j \leq L} }
\vert \sum_{\lambda \in {\bf \Lambda}_{i}} \frac{1}{N_p} 
{\rm e}^{-j \frac{2\pi}{N}\lambda(i-j)} \vert,
\end{equation}
where the DFT submatrix ${{\bf W}_{\Lambda_{i}}}$ is formulated as
\begin{equation}
{{\bf W}_{\Lambda_{i}}} = 
\begin{bmatrix}
1 & {\rm e}^{-j\frac{2\pi}{N}{\lambda_{i_T,1}}} & \cdots 
& {\rm e}^{-j\frac{2\pi}{N}{\lambda_{i_T,1}}(L-1)} \\ 
1 & {\rm e}^{-j\frac{2\pi}{N}{\lambda_{i_T,2}}} & \cdots
& {\rm e}^{-j\frac{2\pi}{N}{\lambda_{i_T,2}}(L-1)} \\ 
\vdots & \vdots & \ddots & \vdots \\ 
1 & {\rm e}^{-j\frac{2\pi}{N}{\lambda_{i_T,N_p}}} & \cdots 
& {\rm e}^{-j\frac{2\pi}{N}{\lambda_{i_T,N_p}}(L-1)}.
\end{bmatrix}
\label{W_Mat}
\end{equation}
According to the periodic structure of the DFT submatrix 
$({{\bf W}_{\Lambda_{i}}})$
and \eqref{W_Mat}, \eqref{Eq11} can be simplified as
\begin{equation}
\mu_{{\bf W}_{\Lambda_{i}}} = \frac{1}{N_p}
\max_{ 1 \leq r \leq L-1}
\Big\vert \sum_{\lambda \in {\bf \Lambda}_{i}}
{\rm e}^{-j \frac{2\pi}{N}\lambda r} \Big\vert.
\label{Eq12}
\end{equation}
By assuming 
$ w = {\rm e}^{-j 2\pi / N} $ and using \eqref{Eq12} and the fact that the optimum set of pilots
(${\bf \Lambda}_{opt}$)
which minimizes the maximum of all coherences, also minimizes the maximum of all coherences to the power of two, the optimum set of pilots will be derived from 
\begin{align}
{\bf \Lambda}_{opt} =\argmin_{ {\bf \Lambda_i} }
\max_{ 1 \leq r \leq L-1}
\Big\vert \sum_{\lambda \in {\bf \Lambda}_{i}} w^{r\lambda} \Big\vert^2.
\label{Eq15}
\end{align}

Inspired by \cite{Pakrooh_OFDM}, we define the CDS of 
${\bf \Lambda}_{i}$
as
${\bf D}_{i} = \{ \lambda_{i,l} - \lambda_{i,k} \vert 
1 \leq l,k \leq N_p ; l \neq k \}$.
By denoting the number of repetitions of a number
$0 \leq d \leq N-1 $ which is a member of the set ${\bf D}_{i}$ as
$\alpha_{d,i_T}$,
the equation \eqref{Eq15} will be simplified as
\begin{align}
\Big\vert \sum_{\lambda \in {\bf \Lambda}_{i_T}} w^{r\lambda} \Big\vert^2 
&= \sum_{d=0}^{N-1} \alpha_{d,i_T} w^{rd}
\\
&= N_p + \sum_{d=1}^{N-1} \alpha_{d,i_T} w^{rd}.
\label{Eq17}
\end{align}

According to the definition of set 
${\bf D}_{i}$,
the number of 0s in the set is $N_p$ which can be expressed as
\begin{equation}
\alpha_{0,i} = N_p, 1\leq i \leq N.
\end{equation}
The cardinality of the set ${\bf D}_{i}$ is equal to $N_p^2$ since 
\begin{equation}
\sum_{d=0}^{N-1} \alpha_{d,i} = N_p^2, 1\leq i \leq N.
\end{equation}
Therefore, it can be concluded that
\begin{equation}
\sum_{d=1}^{N-1} \alpha_{d,i} = N_p ( N_p -1), 1\leq i \leq N.
\end{equation}

According to equation \eqref{Eq17}, the value of
$\vert \sum_{\lambda \in {\bf \Lambda}_{i}} w^{r\lambda} \vert^2$
depends on $r$, $i$, and $\{ \alpha_{d,i} \}$ which is equivalent to the pilot positions.
By defining
\begin{equation}
g(r,i,{\bf \Lambda}) = 
N_p + \sum_{d=1}^{N-1} \alpha_{d,i} w^{rd} 
\label{Eq18}
\end{equation}
and due to the fact that $r$ and $i$ are discrete variables, it can be concluded that there exists a lower bound for
$\max_{ 1 \leq r \leq L-1}
g(r,i,{\bf \Lambda})$
which is computed using the fact that the term is always more than its mean value
which can be expressed as
\begin{equation} \max_{ 1 \leq r \leq L-1} g(r,i_T,{\bf \Lambda}) \geq
\frac{N_p(N-N_p)}{N-1}.
\label{Eq20}
\end{equation}
The problem of finding optimal pilot positions is the minimization of 
$ \max_{ 1 \leq r \leq L-1} g(r,i,{\bf \Lambda})$
and can be solved when the lower bound explained in \eqref{Eq20} is achieved.
It is obvious that the equality is held when
\begin{equation}
g(1,i,{\bf \Lambda})= \cdots = g(L-1,i,{\bf \Lambda})=\frac{N_p(N-N_p)}{N-1},
\label{MustBeTrue}
\end{equation}
which results in
\begin{equation}
\alpha_{1,1}=\cdots=\alpha_{N-1,1}=N_p.
\label{Eq21}
\end{equation}
Since the total number of subcarriers and pilots are respectively equal to $N$ and $N_p$, equation \eqref{Eq21} states that the best choice of pilot positions in the described OFDM system is achieved when the set of pilot positions form a CDS.

A $(\lambda,v,k)$ CDS exists if and only if 
$k^2-k = (v-1) \lambda$.
Having a set of parameters, different CDS with the same parameters can be produced using addition and multiplication of the elements with a constant number.
Having a sample CDS, other CDS with same $(\lambda,v,k)$ parameters can be produced which can be used as pilot positions of different CUEs.

\section{Simulation Results}
\label{SimRes}
In this section, the simulation results are reported. The parameters of the evaluated OFDM system are reported in table \ref{OurSystem_Table}.
In order to observe the performance of the proposed channel estimation algorithm, no coding is used in the system.
The total number of subcarriers and pilots are chosen according to a (91,10,1) CDS.
The initial CDS used for the proposed pilot method is
$\{ 1, 3, 7, 8, 19, 22, 32, 55, 64, 72 \}$.
%OurSystem_Table
\begin{table}[!h]
\caption
{Simulated MIMO-OFDM system parameters }
\begin{center}
\begin{tabular}{|c|c|c} 
\hline
Parameter & Specifications \\
\hline 
N & 91\\
\hline
 $\rm N_p$  & 10\\
\hline
 OFDM symbol time & 200$\mu s$ \\
\hline
 Cyclic prefix length&32 \\
\hline
  Total bandwidth & 455 KHz\\
\hline
 Sampling frequency &  910 KHz\\
\hline
  Doppler frequency & 50 Hz\\
\hline
 Number of paths & 4\\
\hline
\end{tabular}
\end{center}
\label{OurSystem_Table}
\end{table}
The channels are independent Rayleigh multipath fading with 4 significant nonzero taps.
The performance of the system is evaluated in terms of normalized MSE  and the BER for a zero-forcing equalizer based on the channel estimation.

\begin{figure}[!t]
\centering
\includegraphics[width=0.48\textwidth]{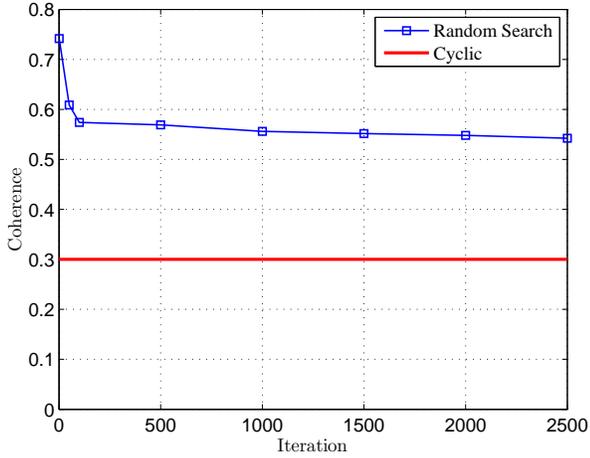}
\vspace{-0.7em}
\caption
{Degradation of coherence vs. number of iterations for the utilized CDS and random search method.}
\label{Fig5}
\end{figure}

Multiple random search method of \cite{Tran_Xueyun_MIMO} is used as a benchmark for our suggested scheme. 
Fig. \ref{Fig5} compares the minimum coherence obtained by the multiple random search \cite{Tran_Xueyun_MIMO} with that of the proposed scheme.
It is observe that the suggested CDS scheme results in a lower coherence while multiple iterations are required for the random search algorithm.

\begin{figure}[!t]
\centering
\includegraphics[width=0.47\textwidth]{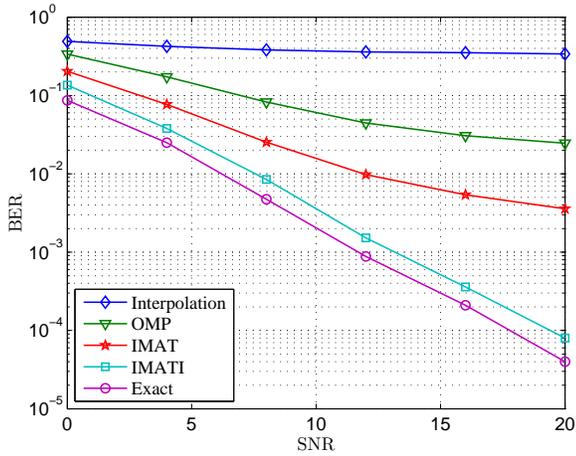}
\vspace{-0.7em}
\caption
{BER vs. SNR of cyclic pilot allocation algorithm (different reconstruction methods)}
\label{Fig3}
\end{figure}

Fig. \ref{Fig3} compares the BER performance of our proposed recovery method with other competing algorithms.
Our sparsity domain smoothing based recovery method is denoted as sparsity domain smoothing IMAT (SDS-IMAT).
The { Interpolation} method is equivalent to a reconstruction method that the whole channel impulse response is constructed by interpolation of the limited random samples corresponding to pilot locations.
The { Exact} method is equivalent to an oracle estimator which knows the channel impulse response perfectly.
The OMP and IMAT algorithms are the proposed schemes of \cite{R002} and \cite{R001}, respectively.
It is shown that our proposed recovery method outperforms other existing methods and has a near-optimal performance.

\begin{figure}[!t]
\centering
\includegraphics[width=0.45\textwidth]{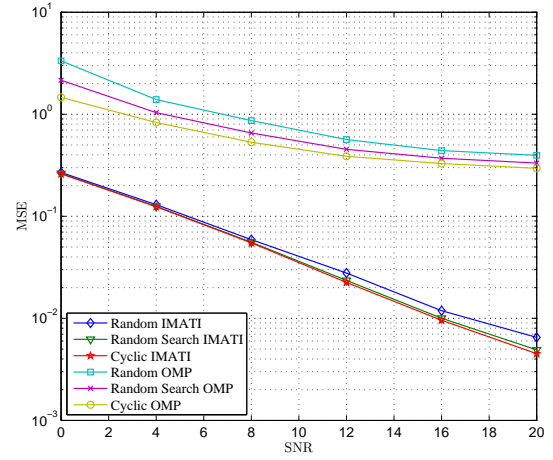}
\vspace{-0.7em}
\caption
{MSE vs. SNR of SDS-IMAT and OMP (different pilot allocation methods)}
\label{Fig2}
\end{figure}

Fig. \ref{Fig2} compares the MSE performance of our proposed SDS-IMAT algorithm with that of the OMP algorithm when random, random search, and cyclic pilot locations are used.
The superiority of our proposed scheme compared to the OMP method for sparse channel estimation in OFDM system is shown in this figure.
According to this figure, we see that our proposed CDS-based pilot allocation scheme has a better performance than the random search and random pilot allocation methods. 

\begin{figure}[!t]
\centering
\includegraphics[width=0.46\textwidth]{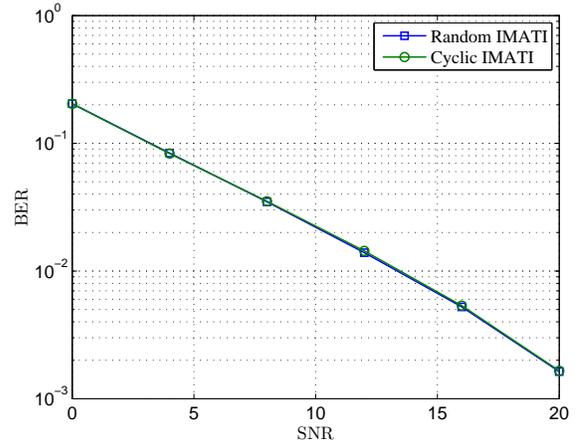}
\vspace{-0.7em}
\caption
{BER of the system vs. SNR for Random and Cyclic SDS-IMAT}
\label{Fig4}
\end{figure}

Fig. \ref{Fig4} demonstrate the effect of having a large number of subcarriers in a OFDM system.
A large size OFDM system with 2257 subcarriers and 48 pilots is assumed. 
We compare our optimized CDS based pilot allocation scheme with random pilot allocation technique, while the SDS-IMAT technique is used as a recovery technique for both. 
The BER versus SNR is depicted in Fig. \ref{Fig4}.
The simulation results show that the performance of these two methods are almost the same in a large size OFDM systems. 
Hence, random pilot allocation scheme combined with our proposed  SDS-IMAT recovery technique can be used in a large size OFDM system to obtain a high quality reconstruction of the channel taps.

\section{Conclusion}
\label{Conc}
In this paper, sparse channel estimation and pilot allocation in an OFDM system has been investigated.
Due to the superior performance of thresholding-based methods, we have modified the IMAT algorithm to present our proposed scheme.
Considering the sparsity domain, we proposed a sparsity domain smoothing-based thresholding recovery method, denoted as SDS-IMAT, which consists of three steps.
The proposed scheme detects nonzero taps of channel impulse response and their corresponding values effectively.
Pilot locations are also found by minimizing the measurement matrix coherence which results in cyclic difference set based pilot locations.
It was shown that the proposed scheme outperforms state-of-the-art methods in terms of BER and MSE.

\bibliographystyle{IEEEtran}
\bibliography{Citations}

\begin{thebibliography}{10}
\providecommand{\url}[1]{#1}
\csname url@samestyle\endcsname
\providecommand{\newblock}{\relax}
\providecommand{\bibinfo}[2]{#2}
\providecommand{\BIBentrySTDinterwordspacing}{\spaceskip=0pt\relax}
\providecommand{\BIBentryALTinterwordstretchfactor}{4}
\providecommand{\BIBentryALTinterwordspacing}{\spaceskip=\fontdimen2\font plus
\BIBentryALTinterwordstretchfactor\fontdimen3\font minus
  \fontdimen4\font\relax}
\providecommand{\BIBforeignlanguage}[2]{{%
\expandafter\ifx\csname l@#1\endcsname\relax
\typeout{** WARNING: IEEEtran.bst: No hyphenation pattern has been}%
\typeout{** loaded for the language `#1'. Using the pattern for}%
\typeout{** the default language instead.}%
\else
\language=\csname l@#1\endcsname
\fi
#2}}
\providecommand{\BIBdecl}{\relax}
\BIBdecl

\bibitem{RMHB_MIMO}
M.~Koolivand, M.~H. Bahonar, and M.~S. Fazel, ``Improving energy efficiency of
  massive {MIMO} relay systems using power bisection allocation for cell-edge
  users,'' in \emph{2019 Iranian Conference on Electrical Engineering
  (ICEE)}.\hskip 1em plus 0.5em minus 0.4em\relax IEEE, 2019, pp. 1470--1475.

\bibitem{RMHB_D2D}
M.~H. Bahonar and M.~J. Omidi, ``Centralized {QoS}-aware resource allocation
  for {D2D} communications with multiple {D2D} pairs in one resource block,''
  in \emph{Electrical Engineering (ICEE), Iranian Conference on}.\hskip 1em
  plus 0.5em minus 0.4em\relax IEEE, 2018, pp. 643--648.

\bibitem{RMHB_Dense}
M.~H. Bahonar and M.~J. Omidi, ``Distributed pricing-based resource allocation for dense
  device-to-device communications in beyond {5G} networks.''\hskip 1em plus
  0.5em minus 0.4em\relax Wiley Online Library, 2021, p. e4250.

\bibitem{RMHB_CV2X}
M.~H. Bahonar, M.~J. Omidi, and H.~Yanikomeroglu, ``Low-complexity resource
  allocation for dense cellular vehicle-to-everything ({C-V2X})
  communications,'' \emph{IEEE Open Journal of the Communications Society}, pp.
  1--1, 2021.

\bibitem{RIRS}
Q.~Wu and R.~Zhang, ``Towards smart and reconfigurable environment: Intelligent
  reflecting surface aided wireless network,'' \emph{IEEE Communications
  Magazine}, vol.~58, no.~1, pp. 106--112, 2019.

\bibitem{Survey_Ref25}
Y.~Li, N.~Seshadri, and S.~Ariyavisitakul, ``Channel estimation for ofdm
  systems with transmitter diversity in mobile wireless channels,'' \emph{IEEE
  Journal on Selected Areas in Communications}, vol. ~17, no.~~3, pp.
  ~461--471, 1999.

\bibitem{R003}
Z.~Qin, J.~Fan, Y.~Liu, Y.~Gao, and G.~Y. Li, ``Sparse representation for
  wireless communications: A compressive sensing approach,'' \emph{IEEE Signal
  Processing Magazine}, vol.~35, no.~3, pp. 40--58, 2018.

\bibitem{R004}
X.~Ma, F.~Yang, S.~Liu, J.~Song, and Z.~Han, ``Sparse channel estimation for
  {MIMO-OFDM} systems in high-mobility situations,'' \emph{IEEE Transactions on
  Vehicular Technology}, vol.~67, no.~7, pp. 6113--6124, 2018.

\bibitem{R005}
C.~Qi, G.~Yue, L.~Wu, Y.~Huang, and A.~Nallanathan, ``Pilot design schemes for
  sparse channel estimation in {OFDM} systems,'' \emph{IEEE Transactions on
  Vehicular Technology}, vol.~64, no.~4, pp. 1493--1505, 2014.

\bibitem{Tran_4}
C.~Qi and L.~Wu, ``A hybrid compressed sensing algorithm for sparse channel
  estimation in {MIMO OFDM} systems,'' in \emph{2011 IEEE International
  Conference on Acoustics, Speech and Signal Processing (ICASSP)}.\hskip 1em
  plus 0.5em minus 0.4em\relax IEEE, 2011, pp. 3488--3491.

\bibitem{Tran_5}
F.~Wan, W.-P. Zhu, and M.~Swamy, ``Semi-blind most significant tap detection
  for sparse channel estimation of {OFDM} systems,'' \emph{IEEE Transactions on
  Circuits and Systems I: Regular Papers}, vol.~57, no.~3, pp. 703--713, 2010.

\bibitem{CS_Donoho}
D.~L. Donoho, ``Compressed sensing,'' \emph{IEEE Transactions on Information
  Theory}, vol. ~52, no.~~4, pp. ~1289--1306, 2006.

\bibitem{Taubock_Leakage_Sparsity_Enhancing}
G.~Taubock, F.~Hlawatsch, D.~Eiwen, and H.~Rauhut, ``Compressive estimation of
  doubly selective channels in multicarrier systems: Leakage effects and
  sparsity-enhancing processing,'' \emph{IEEE Journal of Selected Topics in
  Signal Processing}, vol.~~4, no.~~2, pp. ~255--271, April 2010.

\bibitem{Let_Chen_Pilot_CEO}
J.-C. Chen, C.-K. Wen, and P.~Ting, ``An efficient pilot design scheme for
  sparse channel estimation in {OFDM} systems,'' \emph{IEEE Communications
  Letters}, vol. ~17, no.~~7, pp. ~1352--1355, July 2013.

\bibitem{Tran_12}
X.~He and R.~Song, ``Pilot pattern optimization for compressed sensing based
  sparse channel estimation in ofdm systems,'' in \emph{2010 International
  Conference on Wireless Communications and Signal Processing (WCSP)}.\hskip
  1em plus 0.5em minus 0.4em\relax IEEE, 2010, pp. 1--5.

\bibitem{Let_Chenhao_Deterministic_Pilot}
C.~Qi and L.~Wu, ``A study of deterministic pilot allocation for sparse channel
  estimation in ofdm systems,'' \emph{IEEE Communications Letters}, vol. ~16,
  no.~~5, pp. ~742--744, May 2012.

\bibitem{Pakrooh_OFDM}
P.~Pakrooh, A.~Amini, and F.~Marvasti, ``{OFDM} pilot allocation for sparse
  channel estimation,'' \emph{EURASIP Journal on Advances in Signal
  Processing}, vol. ~2012, no.~~1, pp. ~1--9, 2012.

\bibitem{Tran_Xueyun_MIMO}
X.~He, R.~Song, and W.-P. Zhu, ``Pilot allocation for sparse channel estimation
  in {MIMO-OFDM} systems,'' \emph{IEEE Transactions on Circuits and Systems II:
  Express Briefs}, vol. ~60, no.~~9, pp. ~612--616, Sept 2013.

\bibitem{R006}
L.~Wan, X.~Qiang, L.~Ma, Q.~Song, and G.~Qiao, ``Accurate and efficient path
  delay estimation in {OMP} based sparse channel estimation for {OFDM} with
  equispaced pilots,'' \emph{IEEE Wireless Communications Letters}, vol.~8,
  no.~1, pp. 117--120, 2018.

\bibitem{R001}
F.~Marvasti, M.~Azghani, P.~Imani, P.~Pakrouh, S.~J. Heydari, A.~Golmohammadi,
  A.~Kazerouni, and M.~Khalili, ``Sparse signal processing using iterative
  method with adaptive thresholding ({IMAT}),'' in \emph{2012 19th
  International Conference on Telecommunications (ICT)}.\hskip 1em plus 0.5em
  minus 0.4em\relax IEEE, 2012, pp. 1--6.

\bibitem{MP_Book}
A.~Ben-Israel and T.~N. Greville, \emph{Generalized inverses}.\hskip 1em plus
  0.5em minus 0.4em\relax Springer, 2003, vol. ~13.

\bibitem{CS_Intro}
E.~J. Cand{\`e}s and M.~B. Wakin, ``An introduction to compressive sampling,''
  \emph{IEEE Signal Processing Magazine}, vol. ~25, no.~~2, pp. ~21--30, 2008.

\bibitem{Pakrooh_7}
J.~A. Tropp, ``Greed is good: Algorithmic results for sparse approximation,''
  \emph{IEEE Transactions on Information Theory}, vol. ~50, no. ~10, pp.
  ~2231--2242, 2004.

\bibitem{R002}
J.~A. Tropp and A.~C. Gilbert, ``Signal recovery from random measurements via
  orthogonal matching pursuit,'' \emph{IEEE Transactions on information
  theory}, vol.~53, no.~12, pp. 4655--4666, 2007.

\end{thebibliography}

\end{document}